\documentclass[prc,reprint,nofootinbib,altaffilletter,superscriptaddress,showpacs]{revtex4-1}
\usepackage{amsmath,amssymb,amsfonts}
\usepackage[dvips]{graphicx}
\usepackage{dcolumn}
\usepackage{bm}
\usepackage{bbm}
\usepackage{xcolor}
\usepackage{hyperref}
\usepackage{lineno}
\usepackage{ulem}

\def\figwidth{0.45\textwidth}

\newcommand{\etal}{\textit{et al.}}
\newcommand{\bibpr}[3]{Phys.\ Rev.\ {\bf #1}, #2 (#3)}
\newcommand{\bibprl}[3]{Phys.\ Rev.\ Lett.\ {\bf #1}, #2 (#3)}
\newcommand{\bibprc}[3]{Phys.\ Rev.\ C {\bf #1}, #2 (#3)}
\newcommand{\bibprd}[3]{Phys.\ Rev.\ D {\bf #1}, #2 (#3)}
\newcommand{\bibplb}[3]{Phys.\ Lett.\ B {\bf #1}, #2 (#3)}
\newcommand{\bibnpa}[3]{Nucl.\ Phys.\ A {\bf #1}, #2 (#3)}
\newcommand{\bibnima}[3]{Nucl.\ Instrum.\ Meth.\ A {\bf #1}, #2 (#3)}
\newcommand{\bibepja}[3]{Eur.\ Phys.\ J.\ A {\bf #1}, #2 (#3)}

\newcommand{\bibjpscp}[3]{JPS Conf.\ Proc.\ {\bf #1}, #2 (#3)}
\newcommand{\bibieeens}[3]{IEEE Trans.\ Nucl.\ Sci.\ {\bf #1}, #2 (#3)}
\newcommand{\bibptep}[3]{Prog.\ Theor.\ Exp.\ Phys.\ {\bf #1}, #2 (#3)}
\newcommand{\bibcpc}[3]{Chin.\ Phys.\ C {\bf #1},  #2 (#3)}
\newcommand{\bibepjt}[3]{Eur.\ Phys.\ J.\ Spec.\ Top.\ {\bf  #1}, #2 (#3)}

\begin{document}
%\linenumbers

\title{Coherent photoproduction of the neutral pion and eta meson on the deuteron
at incident energies below 1.15 GeV}
\author{T.~Ishikawa}
\email[Corresponding author: ]{ishikawa@lns.tohoku.ac.jp}
\affiliation{Research Center for Electron Photon Science (ELPH), Tohoku University, Sendai 982-0826, Japan}
\author{A.~Fix}
\affiliation{Tomsk Polytechnic University, Tomsk 634050, Russia}
\author{H.~Fujimura}
\altaffiliation[Present address: ]{Department of Physics, School of Medicine, Wakayama Medical University, Wakayama 641-8509, Japan}
\affiliation{Research Center for Electron Photon Science (ELPH),
Tohoku University, Sendai 982-0826, Japan}
\author{H.~Fukasawa}
\affiliation{Research Center for Electron Photon Science (ELPH),
Tohoku University, Sendai 982-0826, Japan}
\author{R.~Hashimoto}
\altaffiliation[Present address: ]{Institute of Materials Structure Science (IMSS), KEK, Tsukuba 305-0801, Japan}
\affiliation{Research Center for Electron Photon Science (ELPH),
Tohoku University, Sendai 982-0826, Japan}
\author{Q.~He}
\altaffiliation[Present address: ]{Department of Nuclear Science and Engineering, Nanjing University of Aeronautics and Astronautics (NUAA), Nanjing 210016, China}
\affiliation{Research Center for Electron Photon Science (ELPH),
Tohoku University, Sendai 982-0826, Japan}
\author{Y.~Honda}
\affiliation{Research Center for Electron Photon Science (ELPH),
Tohoku University, Sendai 982-0826, Japan}
\author{T.~Iwata}
\affiliation{Department of Science, Faculty of Science, Yamagata University, Yamagata 990-8560, Japan}
\author{S.~Kaida}
\affiliation{Research Center for Electron Photon Science (ELPH),
Tohoku University, Sendai 982-0826, Japan}
\author{J.~Kasagi}
\affiliation{Research Center for Electron Photon Science (ELPH),
Tohoku University, Sendai 982-0826, Japan}
\author{A.~Kawano}
\affiliation{Department of Information Science, Faculty of Liberal Arts, Tohoku Gakuin University, Sendai 981-3193, Japan}
\author{S.~Kuwasaki}
\affiliation{Research Center for Electron Photon Science (ELPH),
Tohoku University, Sendai 982-0826, Japan}
\author{K.~Maeda}
\affiliation{Department of Physics, Graduate School of Science, Tohoku University, Sendai 980-8578, Japan}
\author{S.~Masumoto}
\affiliation{Department of Physics, Graduate School of Science, University of Tokyo, Tokyo 113-0033, Japan}
\author{M.~Miyabe}
\affiliation{Research Center for Electron Photon Science (ELPH),
Tohoku University, Sendai 982-0826, Japan}
\author{F.~Miyahara}
\altaffiliation[Present address: ]{Accelerator Laboratory, KEK, Tsukuba 305-0801, Japan}
\affiliation{Research Center for Electron Photon Science (ELPH),
Tohoku University, Sendai 982-0826, Japan}
\author{K.~Mochizuki}
\affiliation{Research Center for Electron Photon Science (ELPH),
Tohoku University, Sendai 982-0826, Japan}
\author{N.~Muramatsu}
\affiliation{Research Center for Electron Photon Science (ELPH),
Tohoku University, Sendai 982-0826, Japan}
\author{A.~Nakamura}
\affiliation{Research Center for Electron Photon Science (ELPH),
Tohoku University, Sendai 982-0826, Japan}
\author{K.~Nawa}
\affiliation{Research Center for Electron Photon Science (ELPH),
Tohoku University, Sendai 982-0826, Japan}
\author{Y.~Obara}
\affiliation{Department of Physics, Graduate School of Science, University of Tokyo, Tokyo 113-0033, Japan}
\author{S.~Ogushi}
\affiliation{Research Center for Electron Photon Science (ELPH),
Tohoku University, Sendai 982-0826, Japan}
\author{Y.~Okada}
\affiliation{Research Center for Electron Photon Science (ELPH),
Tohoku University, Sendai 982-0826, Japan}
\author{K.~Okamura}
\affiliation{Research Center for Electron Photon Science (ELPH),
Tohoku University, Sendai 982-0826, Japan}
\author{Y.~Onodera}
\affiliation{Research Center for Electron Photon Science (ELPH),
Tohoku University, Sendai 982-0826, Japan}
\author{K.~Ozawa}
\affiliation{Institute of Particle and Nuclear Studies (IPNS), High Energy Accelerator Research Organization (KEK), Tsukuba 305-0801, Japan}
\author{Y.~Sakamoto}
\affiliation{Department of Information Science, Faculty of Liberal Arts, Tohoku Gakuin University, Sendai 981-3193, Japan}
\author{M.~Sato}
\affiliation{Research Center for Electron Photon Science (ELPH),
Tohoku University, Sendai 982-0826, Japan}
\author{H.~Shimizu}
\affiliation{Research Center for Electron Photon Science (ELPH),
Tohoku University, Sendai 982-0826, Japan}
\author{H.~Sugai}
\altaffiliation[Present address: ]{Gunma University Initiative for Advanced Research (GIAR), Maebashi 371-8511, Japan}
\affiliation{Research Center for Electron Photon Science (ELPH),
Tohoku University, Sendai 982-0826, Japan}
\author{K.~Suzuki}
\altaffiliation[Present address: ]{The Wakasa Wan Energy Research Center, Tsuruga 914-0192, Japan}
\affiliation{Research Center for Electron Photon Science (ELPH),
Tohoku University, Sendai 982-0826, Japan}
\author{Y.~Tajima}
\affiliation{Institute of Arts and Sciences, Yamagata University, Yamagata 990-8560, Japan}
\author{S.~Takahashi}
\affiliation{Research Center for Electron Photon Science (ELPH),
Tohoku University, Sendai 982-0826, Japan}
\author{Y.~Taniguchi}
\affiliation{Research Center for Electron Photon Science (ELPH),
Tohoku University, Sendai 982-0826, Japan}
\author{Y.~Tsuchikawa}
\altaffiliation[Present address: ]{J-PARC Center, Japan Atomic Energy Agency (JAEA), Tokai 319-1195, Japan}
\affiliation{Research Center for Electron Photon Science (ELPH),
Tohoku University, Sendai 982-0826, Japan}
\author{H.~Yamazaki}
\altaffiliation[Present address: ]{Radiation Science Center, KEK, Tokai 319-1195, Japan}
\affiliation{Research Center for Electron Photon Science (ELPH),
Tohoku University, Sendai 982-0826, Japan}
\author{R.~Yamazaki}
\affiliation{Research Center for Electron Photon Science (ELPH),
Tohoku University, Sendai 982-0826, Japan}
\author{H.Y.~Yoshida}
\affiliation{Institute of Arts and Sciences, Yamagata University, Yamagata 990-8560, Japan}
\begin{abstract}
Coherent photoproduction of the neutral pion and eta meson on the deuteron,
$\gamma{d}${$\to$}$\pi^0\eta{d}$,
has been experimentally studied
at incident photon energies ranging from the reaction threshold to 1.15 GeV.
The total cross section
demonstrates a rapid rise below 1 GeV.
The data are underestimated by
the existing theoretical calculations
based on quasi-free $\pi^0 \eta$ photoproduction on the nucleon followed by deuteron coalescence.
At the same time,
the data are rather well reproduced by the calculations taking into account
the final-state interaction.
We have also measured for the first time the differential cross sections:
the $\pi^0 \eta$ invariant-mass distribution $d\sigma/dM_{\pi \eta}$,
the $\pi^0 d$ invariant-mass distribution $d\sigma/dM_{\pi d}$,
the $\eta d$ invariant-mass distribution $d\sigma/dM_{\eta d}$,
and
the
distribution over the deuteron emission angle $d\sigma/d\Omega_d$
in the overall center-of-mass frame.
The measured cross section $d\sigma/d\Omega_d$ does not exhibit
strongly backward-peaking behavior predicted by the calculations.
At all incident energies, an increase in $d\sigma/dM_{\eta d}$ near the $\eta d$ threshold is observed, which indicates a bound or virtual $\eta d$ state resulting from a  strong attraction between $\eta$ and a deuteron.
The possibilities of using coherent $\pi^0\eta$ photoproduction on a
nucleus to study the
$\eta$-nuclear
interaction are also discussed.
\end{abstract}

% 13.60.Le Meson production
% 14.40.Be Light mesons (S=C=B=0)
% 25.20.Lj Photoproduction reactions

\maketitle
%\tableofcontents
\section{Introduction}
Meson-nucleus interactions provide crucial insight into
quantum chromodynamics (QCD) as a fundamental theory of strong interaction
in the non-perturbative regime,
where a meson is considered as
an excitation of the QCD vacuum described by various non-vanishing condensates.
The properties of a meson may change in a nucleus
due to
the partial restoration of chiral symmetry,
leading to a decrease of
the chiral condensate~\cite{bern88,brow92,hats92}.
Among mesons, the eta meson ($\eta$) is one of the interesting particles
because of $\eta$-$\eta'$ mixing~\cite{bass06,hire14,bass19}
and coupling with a nucleon ($N$) to the
nucleon resonance $N(1535)1/2^-$, which is a candidate for the chiral partner of $N$~\cite{
deta89,hats89,
jido02,naga03,naga05,jido08,
yori00,kino06}.
A traditional tool for studying the $\eta$-nuclear interaction
is the single $\eta$ production from a nucleus.
A significant increase in the $\eta$ yield at low relative
$\eta$-nuclear momenta
observed in the reactions
$pd\to \eta\,{}^3{\rm He}$~\cite{maye96},
$dp\to \eta\,{}^3{\rm He}$~\cite{smyr07,mers07},
$\gamma \,{}^3{\rm He}\to \eta \,{}^3{\rm He}$~\cite{pfei04,pfei05,pher12},
and
$\gamma \,{}^7{\rm Li}\to \eta \,{}^7{\rm Li}$~\cite{krus14,magh13}
may be interpreted as a signature of attractive forces between
$\eta$ and the nucleus.

A significant amount of information on the  low-energy $\eta$-nuclear
dynamics
has been obtained from
the reactions
$pn\to \eta d$~\cite{cale97,cale98} and $pd\to \eta pd$~\cite{hibo00,bilg04}.
Despite large cross sections of these hadronic processes,
their analysis can be complicated by various ambiguities associated with the initial-state interaction as well as with
dominance of various two-step mechanisms,
leading to undesirable model dependence~\cite{teng05}.

These disadvantages can be overcome by turning to electromagnetic processes.
At first, the electromagnetic interaction may be included perturbatively
up to the first order in $\alpha_{\rm em}$,
so that it is not necessary to take into account the initial-state interaction.
The coherent photoproduction of $\pi^0\eta$ pairs on a nucleus,
$\gamma A\to \pi^0 \eta A$, considered in rather detail in Ref.~\cite{egor13} is especially
suitable for studying the
$\eta$-nuclear
interaction under controlled conditions.
Although there are three hadrons in the final state
of such reactions,
the
$\pi^0$-nuclear
interaction is expected to result in a trivial decrease
of the $\pi^0$ yield due to absorption
by the residual nucleus.
This effect is well-described on a phenomenological level,  and it is thus
well under control.
In addition, the $\pi^0 \eta$ interaction is small at least at low energies
below $a_0(980)$~\cite{acha10}.
Furthermore, in these reactions,
the major fraction of the transferred momentum is carried away by
$\pi^0$ because of its small mass.
As a result, an essential part of the available
kinematic region corresponds to the low relative momentum  between $\eta$ and the nucleus,
i.e., events in which the interaction between these particles is of particular importance.

An additional advantage of $\gamma A\to \pi^0\eta A$ is that the underlying
elementary process $\gamma N\to \pi^0\eta N$
is rather well understood. According to various studies~\cite{doer06,fix10,
naka06,ajak08,
horn08a,horn08b,gutz08,jaeg09,gutz10,gutz14,sokh18,
kash09,kash10,anna15,kaes15,kaes16,kaes18,
krus11,krus15},
the dominant mechanism of $\pi^0\eta$ photoproduction
is the excitation of  the $\Delta(1700)3/2^-$ and $\Delta(1940)3/2^-$ resonances,
followed by their decay into the $\eta\Delta(1232)3/2^+$, and successively
into the final $\pi^0\eta N$ state.
The dominance
of the isospin $I=3/2$  part
leads to an approximate equality of the elementary
amplitudes on the proton and neutron, being in good agreement with the experimental data~\cite{kaes16}.
Due to the isoscalar nature of the deuteron,
the proton and the neutron amplitudes
are added coherently in $\gamma d\to \pi^0\eta d$.
This relatively simple picture distinguishes
$\gamma d\to\pi^0\eta d$
from the hadronic processes such as $pd\to\eta p d$, where
the main reaction mechanism is still not very well understood.

In this paper, we study the $\gamma{d}${$\to$}$\pi^0\eta{d}$ reaction,
in oder to better understand the mechanism for coherent
$\pi^0 \eta$ photoproduction on a nucleus.
We have substantially extended our study from Ref.~\cite{ishi21},
systematically analyzed the measured distributions,
including additional ones, by comparing them with theoretical calculations.
The organization of the paper is as follows.
Our photoproduction experiment using
a detector based on electromagnetic calorimeters
is described in Sec.~\ref{sec:exp}.
The analysis for selecting the $\gamma d \to \pi^0\eta{d}$ events is discussed in Sec.~\ref{sec:ana}.
In Sec.~\ref{sec:res}, presented are the results for the total and differential cross sections.
We determine the low-energy $\eta d $ scattering parameters
from the phenomenological analysis of the results in Sec.~\ref{sec:dis}.
Finally, we summarize our results in Sec.~\ref{sec:sum}.

\section{Experiment}\label{sec:exp}
A series of meson production experiments~\cite{ishi16} were conducted using a bremsstrahlung-photon beam~\cite{ishi10} from
a primary 1.20-GeV electron beam in a synchrotron called the STretcher Booster
(STB) ring~\cite{hino05} at the Research Center for Electron Photon Science (ELPH)~\cite{hama20},
Tohoku University, Japan.
The photon beam was provided by inserting a carbon fiber (radiator)~\cite{obar19} into the stored
electrons in the STB ring,
and then collimated with two lead apertures of 10 and 25 mm in diameter
located at 4.2- and 12.9-m downstream from the radiator, respectively.
A cycle of providing a photon beam in the range from 12 to 26~s
involved the following stages:
\begin{enumerate}
\item injection of 0.15-GeV electrons from a linear accelerator,
\item acceleration of the electrons up to 1.20 GeV,
\item production of bremsstrahlung photons by the radiator, and
\item preparation for the next cycle.
\end{enumerate}
In the third stage, the radiator gradually moved towards the center of the circulating electrons so as to
produce a constant-intensity photon beam. The durations of this stage
were different (6--20 s)
for different experimental periods.
The typical photon intensity was 360~MHz
at normal operation (for the shortest cycle).

To monitor the position and size of the photon beam,
the two-dimensional intensity map at the target location was
 regularly measured~\cite{ishi10,ishi16a}.
The photon-beam size was $\approx 6.5$ and $\approx 7.5$ mm
in the standard deviation for
 the $x$ (horizontal) and $y$ (vertical) directions,
respectively.
The photon energy was determined by detecting the post-bremsstrahlung electron with a
photon-tagging counter, STB-Tagger II~\cite{ishi10}.
The typical photon-tagging rate was 20~MHz corresponding to the photon intensity of 360 MHz, and the tagging energy of the photon beam ranged from 0.75 to 1.15 GeV.
The tagging signal did not always correspond to a photon arrival at the target
since the electron from M{\o}ller scattering or Coulomb multiple scattering at the radiator
may come to the tagging counter,
or the photon may disappear by converting into a positron-and-electron pair even if the corresponding post-bremsstrahlung electron was correctly detected.
Moreover, the electrons entering the tagging counter may
be electrons produced by scattering or reaction with the residual gas in the STB ring.
Thus, the photon transmittance (so called tagging efficiency),
which is the probability for the photon to reach the target
when the electron is detected by the tagging
detector, was also regularly measured~\cite{ishi10,mats18}.
The average of the photon transmittance was $\approx${53\%}
(different for different photon-tagging channels~\cite{ishi10},
and almost constant for different experimental periods in each channel).

The target used in the present experiment was liquid deuterium stored in a cryostat,
which consisted of a refrigerator, a heat transfer pipe including a target cell, and a heat shield.
The refrigerator was a two-stage Gifford-McMahon (GM) cryocooler, Sumitomo Heavy Industries RKD-415D.
It was vertically mounted at 920-mm upstream of the target center
and attached to the heat transfer pipe leading to the target center
because the target was surrounded with detector modules except for the beam hole.
The pipe was made of 99.99\% pure aluminum to obtain high thermal conductivity.
The length of the pipe was 1,000 mm and the cell was located at its end.
The thickness of the cell was 40 mm, and the inner and outer diameters measure 61 and 65 mm, respectively.
The cell was separated from vacuum with the entrance and exit windows made of aramid foil with a thickness of 12.5 $\mu$m.
Since each window was thin and deformed when liquid deuterium was filled into the cell,
the displacements of the window at several positions were measured with a laser displacement meter,
KEYENCE LK-H155, by means of a laser triangulation method~\cite{yama14}.
The effective thickness of the deuterium target was found to be 45.9 mm
from the measured displacements and the intensity map of the photon beam.
Details of the liquid deuterium target can be found elsewhere~\cite{hash08,ishi16b}.

\begin{figure}[t]
\begin{center}
\includegraphics[width=\figwidth]{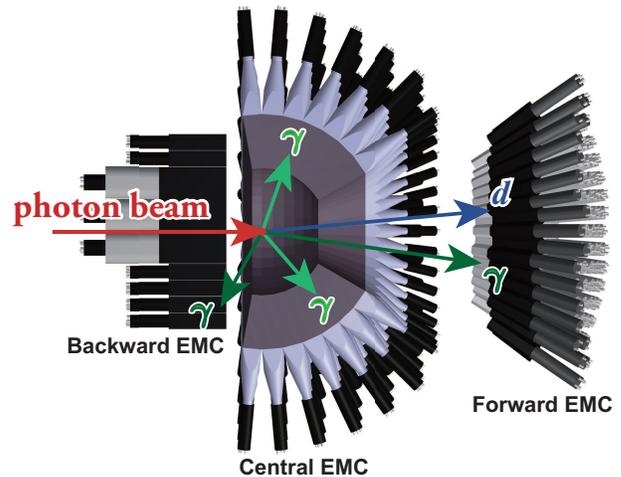}
\end{center}
\caption{Cross sectional view of FOREST.
It consists of three EMCs:
the forward EMC consists of 192 pure CsI crystals,
the central is comprised of 252 lead scintillating-fiber sandwich modules,
and the backward has 62 Cherenkov counters made up of lead glass.
The energy resolution of each EMC is approximately 3\% (forward), 7\% (central), and 5\% (backward) for 1-GeV photons.
A PSH (omitted in the figure) is placed in front of each EMC to identify charged particles.
}
\label{fig01}
\end{figure}

All the final-state particles in the $\gamma{d}${$\to$}$\pi^0\eta{d}$ reaction
were measured with the FOREST detector~\cite{ishi16b}
depicted
in Fig.~\ref{fig01}.
FOREST consisted of three different electromagnetic calorimeters (EMCs):
192 pure cesium-iodide (CsI) crystals for the forward EMC,
252 lead scintillating-fiber sandwich modules for the central, and
62 lead-glass Cherenkov counters for the backward.
The energy responses of the prototype EMCs were investigated using 100--800 MeV positron beams for testing detectors at ELPH~\cite{ishi12},
and the energy resolutions were found to be approximately 3\%, 7\%, and 5\% in response to 1-GeV photons
for the forward, central, and backward EMCs, respectively~\cite{ishi16b}.
A plastic-scintillator hodoscope (PSH) was placed in front of each EMC
to identify charged particles.
The forward PSH consisted of left- and right-handed spiral-shaped plastic scintillators,
and it could determine the impact position of a charged particle even when no corresponding EMC module was found.
FOREST covered the solid angle of 88\% in total.

The digitized data of the energies and timings
for signals from STB-Tagger II and FOREST
were acquired with a network-based data acquisition (DAQ) system
developed for the FOREST experiments~\cite{ishi16b,fuji08}.
The DAQ system consisted of five collector subsystems,
an event-builder subsystem, and a recorder subsystem.
The collector subsystem read
out the digitized data from the front-end electronics
 in parallel with the other collector subsystems,
and converted the data into a single data fragment.
The event builder gathered the fragments from all the collectors, and
constructed a physics event.
The recorder stored the events in a hard disk.
The trigger condition of the DAQ system for processing an event
was made for detecting more than one particle in coincidence with a
photon-tagging signal~\cite{ishi16b}, was the same as that
in Refs.~\cite{ishi19,ishi17,ishi20}.
The trigger rate was $1.7$~kHz on average,
and the DAQ efficiency was $79\%$ on average~\cite{ishi19}.
The details of the DAQ system  can be found elsewhere~\cite{fuji08}.

Regarding the deuterium target at the circulating electron energy of 1.20 GeV,
the number of the events acquired was $2.3\times 10^9$ in total,
and that of the corresponding tagging signals was  $2.0\times 10^{13}$.
Additionally,
we took $2.1\times 10^9$ and  $2.4\times 10^8$ events,
corresponding to $3.1\times 10^{13}$ and $9.7\times 10^{12}$ tagging signals,
for the hydrogen and empty targets at the same electron energy, respectively.
The data acquired with the hydrogen target were mainly used for
checking consistency of geometries, resolutions, and efficiencies of the detector elements
in FOREST.
Those with the empty target were used to confirm that the non-target contributions,
mainly from the two sheets of 12.5~$\mu$m-thick aramid foil,
are negligibly small.
The details of the collected data can be found elsewhere~\cite{ishi11}.

\section{Event selection}\label{sec:ana}

When a photon entered a module in an EMC, not only this module but also
the neighboring modules had a finite energy generated by the electromagnetic (EM) shower.
To obtain the energy and incident position of the photon,
we grouped a set of modules as an EMC cluster
to which the particles in an EM shower deposited energies.
The energy of the EMC cluster was basically given by the sum of the energies of the cluster members.
A minor correction was
made to take into account the energy leakage from the EMC modules.
The incident position was reconstructed by a rational function using the energies of the
maximum-energy module and the adjacent modules.
This position-reconstruction method was studied
using the data obtained in
the beam tests for the prototype EMCs
where positrons were perpendicularly incident on the front face of a module.
Thus the position of the cluster, determined as
an average
of the front-face centers of modules in the cluster
 with energy-dependent weights,
was slightly shifted from the actual incident position
due to the finite injection angle for the forward and backward EMCs.
It should be noted that
the front face of each module in the central EMC is orthogonal
to the line extending from the target center.
The corrections due to the mentioned shift was determined by considering the depth
at the maximum energy deposit of the EM shower.
The energy calibration of each EMC module was performed
in such a way that the $\pi^0$ peak in the $\gamma\gamma$ invariant-mass distribution corresponds to the $\pi^0$ mass.
Details of the energy and position reconstruction, including energy calibration, are described
in Ref.~\cite{ishi16b}

The event selection was carried out
for the $\gamma{d}${$\to$}$\pi^0\eta{d}${$\to$}$\gamma\gamma\gamma\gamma{d}$ reaction.
Initially, events containing four neutral particles and a charged particle
were selected.
The EMC cluster without the corresponding PSH hit was recognized as a neutral particle.
The PSH hit gave a charged particle regardless of the presence of the corresponding EMC cluster.
The time difference between any two of the four neutral EMC clusters
was required to be less than three times
of the time resolution
depending on  the EMCs that the two clusters belonged to
and the energies they deposited.
The events were selected in such a way that a charged particle was detected with the forward PSH
provided that
the time delay from the response of the four neutral clusters was greater than 1 ns,
and the deposit energy of a charged particle in PSH was larger than twice that of the minimum-ionizing particle.

Further selection was made by applying a kinematic fit (KF) with six constraints (6C)
for the $\gamma{d}${$\to$}$\pi^0\eta d$ hypothesis (CKF):
the energy and the three-momentum
are conserved between the initial and final states,
the invariant mass of two of the four photons is equal to the $\pi^0$ mass,
and that of the other two photons is equal to the $\eta$ mass.
In each event, the most probable combination of
dividing the four photons into two groups of
two photons, each group corresponding to the $\pi^0$ and $\eta$ mesons,
was chosen.
The momentum of the charged particle was obtained from the time delay
under the assumption that the charged particle has the deuteron mass.
Only those events were selected whose
$\chi^2$ probability (probability value for resulting $\chi^2$ with the degrees of freedom in a fit)
 was greater than
0.2 in order to reduce the number of events from other background reactions.
The most competitive background contribution came from deuteron misidentification in the quasi-free (QF) reaction $\gamma{p'}${$\to$}$\pi^0\eta p$,
where $p'$ denotes the proton bound in the deuteron.
Thus, an additional criterion for the selected events was the requirement that
the
$\chi^2$ probability
is less than 0.01
in another KF for the $\gamma{p'}${$\to$}$\pi^0\eta p$ hypothesis (QFKF).
In this QFKF, the $x$, $y$, and $z$ momenta of the initial bound proton were
assumed to be measured with a centroid of 0 MeV/$c$ and
a resolution of 40 MeV/$c$,
and its total energy was obtained
assuming that the spectator neutron is always on-shell.
Finally, sideband-background subtraction was performed
for accidental-coincidence events detected in STB-Tagger II and FOREST.

\begin{figure}[t]
\begin{center}
\includegraphics[width=\figwidth]{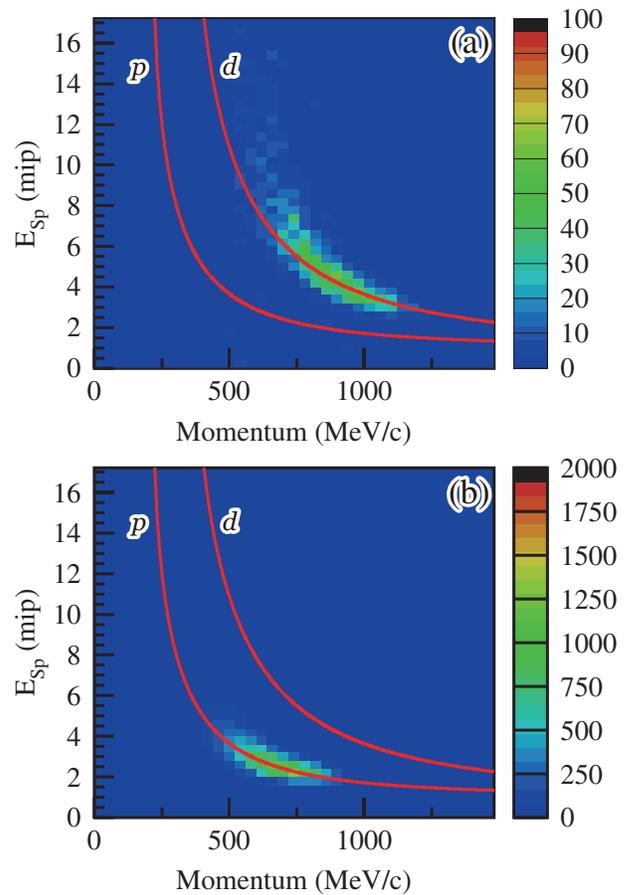}
\end{center}
\caption{Correlations between the measured energy with PSH
and momentum of a charged particle
in the coherent $\gamma d \to \pi^0 \eta d$ reaction (a) and QF $\gamma p' \to \pi^0 \eta p$ reaction (b).
Here the absolute value of the
missing momentum $\vec{p}_X$ is provided
for the $\gamma d\to \pi^0\eta X$  reaction.
The lower and upper curves in each panel show the loci corresponding to correlations for protons ($p$) and
deuterons ($d$), respectively. }
\label{fig02}
\end{figure}

The 6C KF was rather effective for selecting the
$\gamma d\to \pi^0\eta d$ reaction.
Figure~\ref{fig02} shows the correlation between the energy measured with PSH
and the momentum of a charged particle
in the coherent $\gamma d \to \pi^0 \eta d$ reaction and in the QF $\gamma p' \to \pi^0 \eta p$ reaction.
Here, the momentum of the charged particle is determined as the
difference between the
incident-photon momentum and four final-state photon momenta:
\begin{equation}
\vec{p}_X = \vec{p}_0 - \sum_{i=1}^4 \vec{p}_i
\end{equation}
where $\vec{p}_0$ stands for the momentum of the incident photon, and
$\vec{p}_i$
 $(i=1,\ldots, 4)$ denotes that of the $i$-th final-state photon from the $\pi^0$ or $\eta$ decays.
Since the momentum of the target deuteron is zero,
$\vec{p}_X$ equals the momentum of the final-state deuteron.
In Fig.~\ref{fig02}, the events for each reaction were selected by the only requirement
that the $\chi^2$ probability is larger than 0.2 in the corresponding KF.
Here, the additional requirement ($\chi^2$ probability is lower than 0.01 in QFKF)
was not applied for selecting the coherent events.
The deuteron-detected events were correctly selected in CKF,
and the proton-detected events were selected in QFKF.

\begin{figure}[t]
\begin{center}
\includegraphics[width=\figwidth]{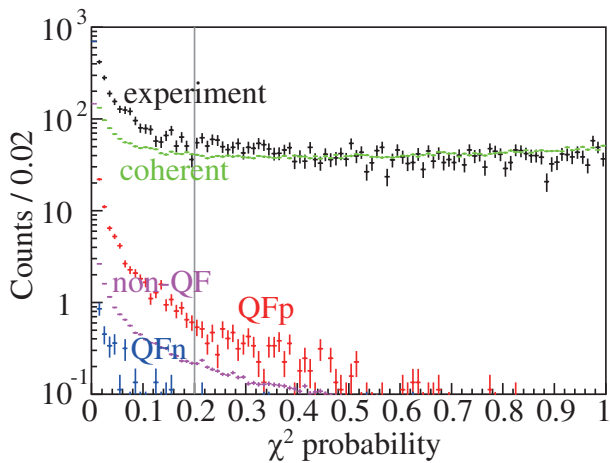}
\end{center}
\caption{$\chi^2$ probability distributions in CKF
for the experimental data,
the coherent $\gamma d\to \pi^0\eta d$ reaction,
and the possible background reactions.
The distributions are plotted for the highest-energy group of photon-tagging channels
divided into four ($E_\gamma=1.01$--1.15 GeV).
The $\chi^2$ probability is lower than 0.01 in QFKF.
The black markers show the distribution for the experimental data,
and the green, red, blue, and magenta
represent the distributions for the
coherent,
QF$p$,
QF$n$,
and non-QF reactions, respectively.
}
\label{fig03}
\end{figure}
Background contamination was taken into account in the event selection  described above
using CKF ($\chi^2$ probability was higher than 0.2) and QFKF ($\chi^2$ probability was lower than 0.01).
Figure~\ref{fig03} shows the experimentally obtained $\chi^2$-probability distribution
in CKF for $E_\gamma=1.01$--1.15 GeV
together with those for the coherent $\gamma d\to\pi^0\eta d$ reaction
 and the possible background reactions.
The QF $\gamma p'\to \pi^0 \eta p$ reaction (QF$p$)
was the most competitive among the background reactions,
having ten times larger total cross section at maximum~\cite{kaes15}.
The $\chi^2$-probability distribution in CKF
was estimated for QF$p$
when the $\chi^2$ probability was less than 0.01 in QFKF.
To incorporate the acceptance of the events in the FOREST detector,
we adopted a Monte-Carlo (MC) simulation code based on Geant4~\cite{geant4a,geant4b,geant4c}.
In this QF$p$ simulation, the Fermi momentum of the initial bound proton was determined
using the Hulth\'{e}n wave function~\cite{whit60}
with parameters $(\alpha,\beta) = (45.6,234)$ MeV which reproduce
the momentum distribution of nucleons in a deuteron derived
from the $d(e,e'p)n$ reaction~\cite{bern81}.
The total energy of the initial proton was evaluated in the same way as in QFKF.
The three-momenta of the final-state particles
were obtained by the pure phase-space generation
at the center-of-mass (CM) energy of the incident photon and initial bound proton $W_{\gamma p'}$.
The lower limit 0.2 in CKF made the QF$p$ contamination less than 0.5\%.

The QF $\gamma n'\to \pi^0\eta n$ reaction (QF$n$) was also considered
as a candidate for the background reactions
since the charged particle could be emitted when the neutron
was passing through the material.
The $\chi^2$ probability distribution for QF$n$ was
estimated similarly to that for QF$p$.
The  QF$n$ contamination was found to be much less than 0.1\%.
The non-quasi-free $\gamma d\to \pi^0\pi^0pn$ reaction (non-QF)
could be a possible background
although the total cross section for non-QF was negligibly small~\cite{kaes15}.
The $\chi^2$ probability distribution was estimated for non-QF assuming the pure phase-space generation.
The non-QF contamination was found to be less than 0.2\% even
if the total cross section is one tenth of the coherent reaction as shown in Fig.~\ref{fig03}.

\section{Cross sections}\label{sec:res}
Here, we deduce the cross sections for $\gamma d\to\pi^0\eta d$.
In Sec.~\ref{sec:acc}, we describe the method used to estimate the acceptance for this reaction.
The total, differential, and double-differential cross sections are presented in
Secs.~\ref{sec:tcs}, \ref{sec:dcs}, and \ref{sec:ddcs}, respectively.

\subsection{Acceptance}\label{sec:acc}
The acceptance of $\gamma\gamma\gamma\gamma d$ detection for the $\gamma d\to \pi^0\eta d$ events depends on the kinematic
condition determined by detector efficiencies, geometrical acceptance, and
event selection in the analysis.
Thus, acceptance was estimated by the MC simulation
code as a function of the kinematic binning for the cross section.
The generated events in the simulation were reconstructed with the same analysis code
which had been used for the experimental data.
The acceptance for a given incident energy $E_\gamma$ and kinematic condition  $\mathcal{K}$,
specified by a set of momentum vectors for all the final-state particles,
 is given by
\begin{equation}
\mathcal{A}\left(E_\gamma,\mathcal{K}\right)
=\frac{
\mathcal{N}_{\rm acc}\left(E_\gamma,\mathcal{K}\right)
}{
\mathcal{N}_{\rm gen}\left(E_\gamma,\mathcal{K}\right)
}\,,
\end{equation}
where $\mathcal{N}_{\rm gen}\left(E_\gamma,\mathcal{K}\right)$
and $\mathcal{N}_{\rm acc}\left(E_\gamma,\mathcal{K}\right)$ denote
the number of the generated events and that of accepted events for $\mathcal{K}$,
respectively.
Since the $\gamma d\to \pi^0\eta d$ reaction has three particles in the final state,
the number of independent variables is five to specify $\mathcal{K}$,
and it is reduced to four assuming rotational invariance with respect to
the photon beam axis.
Also assumed is an additional rotational symmetry
of the plane spanned by the momenta of two final particles
in the overall CM frame
with respect to the momentum of the third particle.
In this regard, we have chosen the following three variables for fixing  $\mathcal{K}$:
the $\pi^0{d}$ invariant mass $M_{\pi{d}}$,
the $\eta{d}$ invariant mass $M_{\eta{d}}$, and
the deuteron emission angle $\theta_d$ in the overall CM frame
with the $z$-axis along the incident photon momentum
(the azimuth $\phi_d$ drops out
due to the rotational symmetry mentioned above).

To determine the total cross section,
we have to obtain the expected value of $\mathcal{A}\left(E_\gamma,\mathcal{K}\right)$
for all possible $\mathcal{K}$s:
\begin{equation}
\mathcal{A}\left(E_\gamma\right)
=
\langle\mathcal{A}\left(E_\gamma,\mathcal{K}\right)
\rangle=
\frac{\displaystyle
\sum_{\mathcal{K}}
\mathcal{N}_{\rm acc}\left(E_\gamma,\mathcal{K}\right)
}{\displaystyle
\sum_{\mathcal{K}}
\mathcal{N}_{\rm gen}\left(E_\gamma,\mathcal{K}\right)}\,.
\end{equation}
Similarly, we have to obtain the expected value
at a fixed $V_i$
to deduce the differential cross section as a function of $V_i$:
\begin{equation}
\mathcal{A}\left(E_\gamma,V_i\right)
=\langle
\mathcal{A}\left(E_\gamma,\mathcal{K}[V_i]\right)
\rangle
=\frac{\displaystyle
\sum_{\mathcal{K}[V_i]}
\mathcal{N}_{\rm acc}\left(E_\gamma,\mathcal{K}\right)
}{\displaystyle
\sum_{\mathcal{K}[V_i]}
\mathcal{N}_{\rm gen}\left(E_\gamma,\mathcal{K}\right)
}
\end{equation}
where it is explicitly denoted that
$\mathcal{K}$ depends on the running variable $V_i$.
It should be emphasized that $V_i$ does not have to be one of
the kinematic variables
$M_{\pi{d}}$,
$M_{\eta{d}}$, or $\cos\theta_d$.
Apparently,
$\mathcal{A}\left(E_\gamma\right)$ and
$\mathcal{A}\left(E_\gamma,V_i\right)$ depend on
$\mathcal{N}_{\rm gen}\left(E_\gamma,\mathcal{K}\right)$.
Thus, the event generation should be adjusted
to reproduce the measured distributions
over $M_{\pi{d}}$, $M_{\eta{d}}$,  and $\cos\theta_d$.

Because of the limited statistics,
each distribution for the accepted events was
compared to the measured one for each of
three higher-energy groups of photon-tagging channels
divided into the three bins
($E_\gamma=0.85$--0.94, 0.95--1.01, 1.01--1.15 GeV).
For comparison,
histograms with a bin width of 10 MeV
for the $M_{\pi{d}}$ and $M_{\eta{d}}$ distributions,
as well as histograms with a bin width of 0.2 for $\cos\theta_d$ were built.
The pure phase-space event generation has been
changed to reproduce all the measured
$M_{\pi{d}}$, $M_{\eta{d}}$,  and $\cos\theta_d$ distributions within the statistical uncertainties.
Here,  a factorized weight was introduced for
providing $\mathcal{N}_{\rm gen}(E_\gamma,\mathcal{K})$ with continuous and
smooth functions: $w_{\pi{d}}\left(E_\gamma,M_{\pi{d}}\right)$,
$w_{\eta{d}}\left(E_\gamma,M_{\eta{d}}\right)$, and
$w_{d}\left(E_\gamma,\cos\theta_d\right)$.

\subsection{Total cross section}\label{sec:tcs}
\begin{figure}[hbt]
\begin{center}
\includegraphics[width=\figwidth]{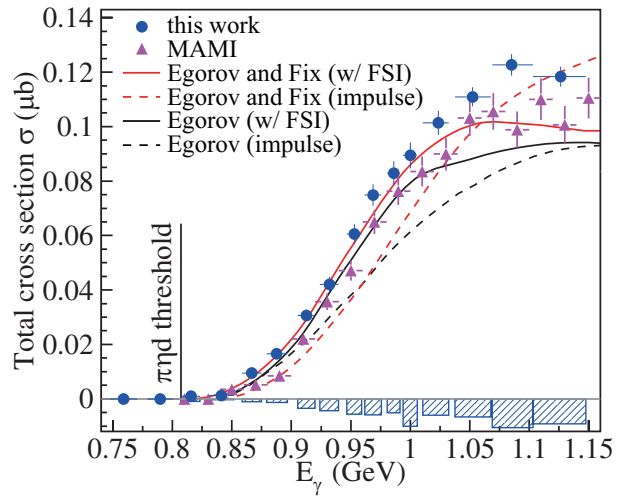}
\end{center}
\caption{Total cross section $\sigma$ as a function of the incident photon energy
$E_\gamma$.
The blue circles show $\sigma$ obtained in this work, and the magenta squares
show that obtained at the MAMI facility~\cite{kaes15}.
The horizontal error of each data point corresponds to
the coverage of the incident photon energy, and the vertical error shows
the statistical error of $\sigma$.
The lower hatched histogram shows the systematic error of $\sigma$ in this work.
The data are compared with the theoretical calculations
given by Egorov and Fix~\cite{egor13} (red curves)
and by Egorov~\cite{egor20} (black curves).
The dashed and solid curves are obtained with
the impulse approximation and with FSI, respectively.
}
\label{fig04}
\end{figure}

The total cross section of the $\gamma d \to \pi^0\eta d$ reaction
was obtained according to the formula:
\begin{equation}
\sigma(E_\gamma) = \frac{
\mathcal{N}_{\pi\eta d}(E_\gamma)
}{
\mathcal{N}'_\gamma\,
\mathcal{N}_\tau\,
\mathcal{A}(E_\gamma)\,
\mathcal{B}({\pi^0\to \gamma\gamma})\,
\mathcal{B}({\eta\to \gamma\gamma})
\label{eq:1}
}\,,\end{equation}
containing
the number of events for the $\gamma d\to\pi^0\eta d$ reaction $\mathcal{N}_{\pi\eta d}(E_\gamma)$,
the effective number of incident photons $\mathcal{N}'_\gamma$,
the number of target deuterons per unit area $\mathcal{N}_\tau=0.237$ b${}^{-1}$
corresponding to the thickness of 45.9 mm,
the acceptance of the final state $\pi^0\eta d\to \gamma\gamma\gamma\gamma d$ detection
$\mathcal{A}(E_\gamma)$, and the branching ratio of the $\pi^0\to\gamma\gamma$ decay
$\mathcal{B}({\pi^0\to \gamma\gamma})=98.823\%$,
and that of the $\eta \to \gamma\gamma$ decay
$\mathcal{B}({\eta \to \gamma\gamma})=39.41\%$~\cite{pdg}.
$\sigma(E_\gamma)$ was obtained for each group of
photon-tagging channels divided into sixteen intervals.
The number of incident photons $\mathcal{N}_\gamma$ was determined by
multiplying the number of tagging signals after the counting-loss correction
by the corresponding photon transmittance.
$\mathcal{N}'_\gamma$ was additionally obtained by multiplying
 $\mathcal{N}_\gamma$ by the DAQ efficiency.
Figure~\ref{fig04} shows the total cross section $\sigma$
as a function of the incident photon energy $E_\gamma$.

We estimated the following systematic uncertainties
pertaining to the $\sigma$ measurement:
yield extraction associated with event selection in KF;
acceptance correction arising from the $M_{\pi{d}}$, $M_{\eta{d}}$,
and $\cos\theta_d$ distributions
in event generation of the simulation,
from the FOREST coverage, and
from the detection efficiency of deuterons;
and normalization resulting from the numbers of target deuterons and
incident photons.
At first, to obtain the systematic uncertainty in KF, we
changed the lower limit of event selection in CKF  from 0.1 to 0.5.
We considered the standard deviation of the numbers of the accepted events,
ranging from 4\% to 6\%, as the uncertainty.
Since all relative uncertainties, except that for the overall normalization,
became larger with decreasing the incident photon energy,
we have taken here
those for the incident photon energy higher than 1~GeV
as typical values.
To obtain the effects of the uncertainties of the assumed distributions
in the event generation, we introduced an additional factor:
\begin{equation}
w_i (E_\gamma,R_i) = 1 + \alpha_i R_i (E_\gamma,V_i)
\end{equation}
where $V_i$ is a variable from the set
$M_{\pi{d}}$,  $M_{\eta{d}}$, and $\cos\theta_d$.
$R_i$ is a real value obtained from $V_i$  by mapping
to the interval [0,1]:
\begin{equation}
R_i(E_\gamma,V_i)= \frac{
V_i-V_i^{\rm min}(E_\gamma)}{
V_i^{\rm max}(E_\gamma)-V_i^{\rm min}(E_\gamma)}\,.
\end{equation}
The variation of the estimated acceptance was taken as an uncertainty
when $\alpha_i$ was varied under
reproducing each measured distribution of three kinds
within the statistical uncertainty.
The uncertainties in the acceptance  were
found to be
0.1\%--0.2\%,
0.2\%--0.4\%, and
0.6\%--0.7\%
for the $M_{\pi{d}}$,  $M_{\eta{d}}$, and $\cos\theta_d$
distributions, respectively.
The uncertainty caused by the limited FOREST acceptance was
0.4\%--1.0\%.
The uncertainty in the deuteron detection efficiency was 2.7\%--4.2\%
due to the uncertainty in the density of the vacuum chamber
surrounding the liquid deuterium target.
The normalization uncertainties resulting from the
number of target deuterons and from the number of incident photons
were 1\% and 2.3\%--2.7\%, respectively.
The total systematic uncertainty $\delta\sigma_{\rm syst}$
shown in Fig.~\ref{fig04}
was obtained by combining all
the uncertainties, $\delta\sigma_{\rm syst}^i$s, described above
in quadrature:
\begin{equation}
\delta\sigma_{\rm syst}^2 = \sum \left(\sigma_{\rm syst}^i\right)^2.
\end{equation}

Our results for the total cross section are presented in Fig.\,\ref{fig04}. The MAMI data~\cite{kaes15} are also shown for comparison. As can be seen, there is a slight systematic excess of our data, about $10\%$ on average, which, however, is within the limits of the systematic error.
Comparison with the theoretical calculations reveals fairly good
reproducibility of the model in Ref.\,\cite{egor13}. It is worth noting that at lower energies the impulse approximation with plane waves (red dashed curve) significantly underestimates the observed cross section. Its dependence on the total
CM energy $W$ is mainly determined by the reaction phase space $\mathcal{P}$, which for our reaction with three particles in the final state may be presented in the form
\begin{equation}\label{eq:1a}
\displaystyle
\mathcal{P}\sim \frac{1}{WE_\gamma}\int_{m_\eta+M_d}^{W-m_\pi}\,q_\pi(M_{\eta d})\,p_\eta(M_{\eta d})\,dM_{\eta d}\,,
\end{equation}
where
\begin{equation}\label{eq:1b}
\begin{array}{l}
\displaystyle
q_\pi=
\frac{1}{2W}\,\lambda^{1/2}\left(W,M_{\eta d},m_\pi\right),\, \vphantom{\frac{\sum}{\sum}}\rm\ and\\
\displaystyle
p_\eta=
\frac{1}{2M_{\eta d}}\,\lambda^{1/2}\left(M_{\eta d},m_\eta,M_d\right)
\vphantom{\frac{\sum}{\sum}}%\nonumber
\end{array}
\end{equation}
are the pion momentum in the overall CM frame, and the $\eta$ momentum in the $\eta d$ CM frame, respectively.

The interaction between the $\eta$ meson and the deuteron, due to its attractive character, leads to a visible increase of the plane-wave cross section. This effect is very typical for the processes in which interaction between the final particles is predominantly attractive.
Within the time-independent quantum scattering theory, this is explained by the fact that the reaction rate is determined by the square of the matrix element sandwiched between the initial and final state vectors, namely
$|\gamma d\rangle$ and $|\pi^0\eta d\rangle$ in our case, respectively. The latter, in turn, is proportional to the probability of finding the final particles in the region in which the initial interaction, leading to particle production, acts. Since attraction obviously tends to keep the final particles in this region, its presence leads to a general increase of the reaction yield. The attractive character of the $\eta d$ interaction in the $S$ wave is especially important at low energies, i.e., where the small values of $p_\eta$ account for a significant fraction of the reaction phase space in Eq.~(\ref{eq:1a}). In our case, this results in strong increase of the plane-wave cross section at $E_\gamma<0.9$ GeV.

As the energy increases, the FSI effect changes sign at
about 1.05 GeV and at $E_\gamma=1.15$ GeV results in a reduction of
about 20\,$\%$ of the total cross section. This effect is expected
to originate from the absorption of $\eta$ mesons caused by their
conversion into pions via $\eta N\to\pi N$. The $\eta$ absorption due to inelastic resonance mechanisms is well known to be especially significant in inclusive processes on heavier nuclei~\cite{Krusche_eta_A}, where it leads to the so-called surface production. In addition to the already mentioned pion absorption, the $\eta$ absorption results in an additional decrease
of the reaction yield over the entire energy range under consideration.
Since the relative fraction of low $\eta d$ energies, at which the $\eta d$ attraction is significant, decreases with increasing photon energy, the role of absorption becomes especially noticeable at higher energies. As is seen from Fig.\,\ref{fig04} this interplay between the enhancement due to $\eta d$ attraction and the absorption effects leads shifts the total cross section towards the threshold region.

In general, as one can conclude from Fig.\,\ref{fig04}, inclusion of the $\eta d$ interaction makes it possible to reproduce the experimental cross section in the energy range up to $E_\gamma<1$ GeV.
Above this region, some deviation between the theoretical calculation and the data is observed.
In addition to the fact that the model of Ref.~\cite{egor13} may somewhat overestimate the role of absorption effects mentioned above, another possible reason of this discrepancy is that this model
describes the elementary
 cross section $\gamma N\to \pi^0\eta N$ only at the laboratory photon energies below 1.4 GeV. At higher energies, the resonances that are not included in the model must come into play, so that the
calculation may visibly underestimate the data. When turning to a nucleus, this energy region is effectively incorporated into the calculation due to the Fermi motion, thus leading to some discrepancy which we just observe at higher energies.

The conclusions above applies also to the calculations presented in
Ref.~\cite{egor20} (black curves in Fig.\,\ref{fig04}). Here the black solid curve shows the full calculation in which the FSI effects are taken into account in a unified microscopic approach, including pion rescattering on the spectator nucleon ($\gamma N_1 \to \pi^\pm\pi^0N_1$ followed by $\pi^\pm N_2 \to \eta N_2$ and $N_1N_2\to d$), as well as absorption of an additionally produced pion on the spectator nucleon ($\gamma N_1 \to \pi^\pm\pi^0\eta N_1$ followed by $\pi^\pm N_2 \to N_2$ and $N_1N_2\to d$).
In this case, the full model is also in general agreement with the data in the region below 1 GeV. However, as the energy increases, the description is getting worse, as in the previous case.

\subsection{Differential cross sections}\label{sec:dcs}

\begin{figure*}
\begin{center}
\includegraphics[width=0.8\textwidth]{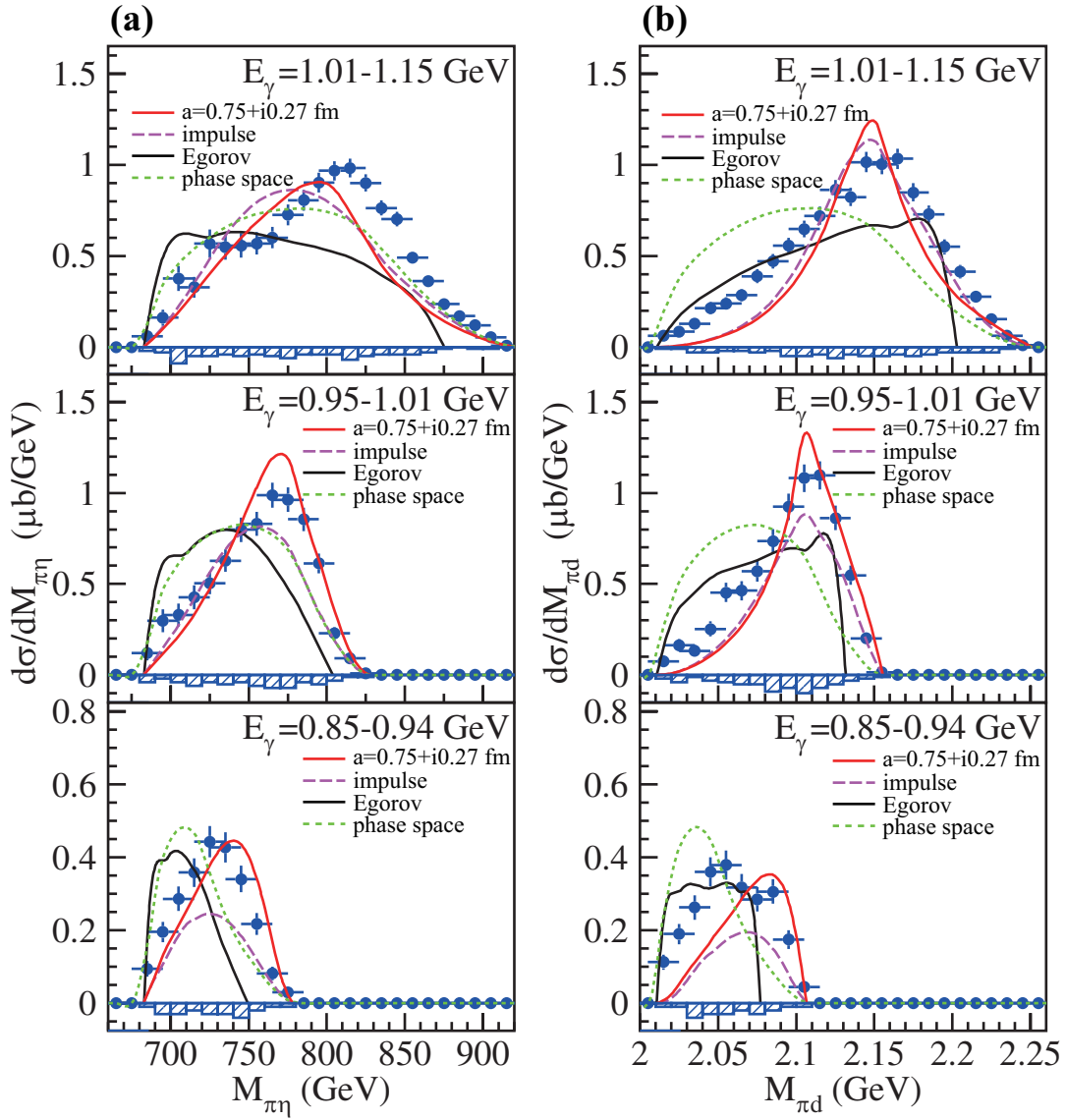}
\end{center}
\caption{Differential cross sections
 $d\sigma/dM_{\pi\eta}$ (a)
and $d\sigma/dM_{\pi d}$ (b)
for $E_\gamma=0.85$--0.94 GeV (bottom), 0.95--1.01 GeV (center),
and 1.01--1.15 GeV (top).
The lower hatched histograms show
the corresponding systematic errors.
The green dotted curves represent the pure phase space.
The magenta dashed, and the red solid curves show
the impulse-approximation calculations, and
the full calculations with the set of the $\eta N$ parameters corresponding to
$a_{\eta N}=0.75+i\,0.27$ fm by Egorov and Fix~\cite{egor13}.
The black solid curves show the full calculations by Egorov~\cite{egor20}
at $E_\gamma=0.90$, 0.98, and 1.08 GeV
in the bottom, center and top panels,
respectively.
The phase space is normalized so that the
corresponding total cross section matches the measured one.
}
\label{fig05}
\end{figure*}

\begin{figure*}
\begin{center}
\includegraphics[width=0.8\textwidth]{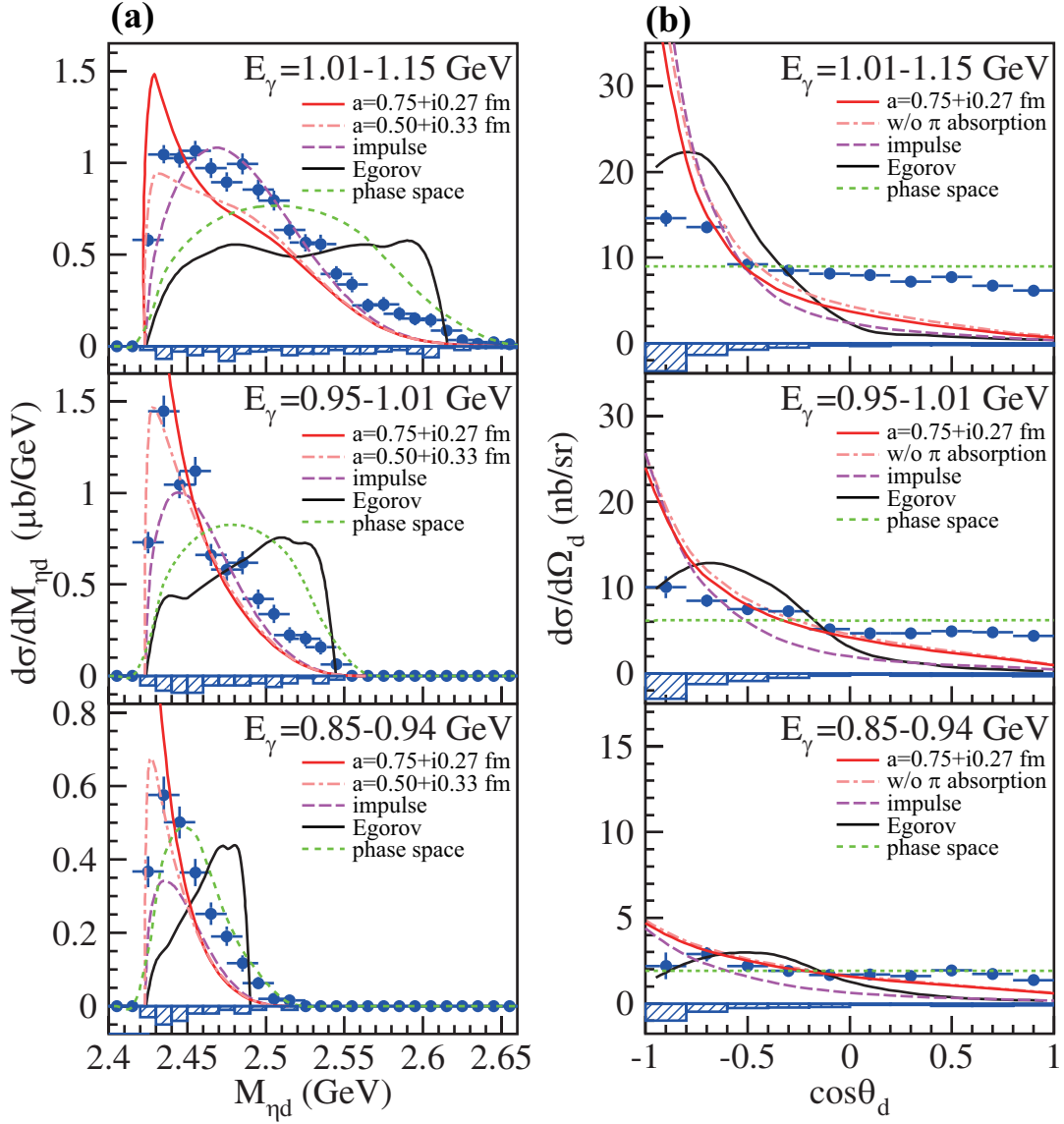}
\end{center}
\caption{Differential cross sections
$d\sigma/dM_{\eta{d}}$ (a) and
$d\sigma/d\Omega_d$ (b).
The meaning of the curves is the same as in Fig.~\ref{fig05},
except for the orange dot-dashed curves which are
obtained with the sets of the $\eta N$ parameters,
corresponding to $a_{\eta N}=0.50+i\,0.33$ fm in $d\sigma/dM_{\eta d}$,
and $a_{\eta N}=0.75+i\,0.27$ fm in $d\sigma/d\Omega_d$.
The pion absorption effect is not included in the latter case,
and that is different from the red solid curves.
}
\label{fig06}
\end{figure*}

To obtain more information about the important mechanisms of the $\gamma d\to\pi^0\eta d$ reaction, we have also measured the invariant-mass and angular distributions:
$d\sigma/dM_{\pi \eta}$,
$d\sigma/dM_{\pi d}$,
$d\sigma/dM_{\eta d}$,
and
$d\sigma/d\Omega_d$ in the energy bins $E_\gamma=0.85$--0.94, 0.95--1.01, and 1.01--1.15 GeV.
The differential cross section
$d\sigma/dV_i$
is given by
\begin{widetext}
\begin{equation}
\frac{d\sigma}{dV_i}(E_\gamma,V_i) = \frac{
\mathcal{N}_{\pi\eta d}(E_\gamma,V_i)
}{
\mathcal{N}'_\gamma\,
\mathcal{N}_\tau\,
\mathcal{A}(E_\gamma,V_i)\,
\mathcal{B}({\pi^0\to \gamma\gamma})\,
\mathcal{B}({\eta\to \gamma\gamma})\,
\Delta V_i
\label{eq:dcs}
}\end{equation}
\end{widetext}
with $V_i$ being one of the quantities $M_{\pi \eta}$,
$M_{\pi d}$, $M_{\eta d}$, or $\cos\theta_d$.
Due to axial symmetry with respect to the incident beam direction
the simple relation $d\sigma/d\Omega_d = d\sigma/d\cos\theta_d/2\pi$ holds.

The results are presented in Figs.\,\ref{fig05} and \ref{fig06}.
The experimental data are depicted by blue circles with statistical errors, whereas the systematic uncertainties are shown by the hatched histograms.
Here, the total systematic uncertainties for the differential cross sections
were obtained by combining all the uncertainties
discussed in Sec.~\ref{sec:tcs}
except those
for the overall normalization (the uncertainties in the number of the target deuterons and in the number of the incident photons).
The data are compared with the pure phase space plotted by the green dotted curves. The magenta dashed, red solid, and orange dotted curves
show the calculations based on the impulse approximation,
the full calculation with FSI from~\cite{egor13} with different sets of the $\eta N$ interaction parameters (corresponding to
$a_{\eta N}=0.75+i\,0.27$ fm~\cite{gree97} and
$a_{\eta N}=0.50+i\,0.33$ fm~\cite{wilk93}). The black solid curves is the full calculations by Egorov~\cite{egor20}.

The mass distribution $d\sigma/dM_{\pi\eta}$ shown in Fig.~\ref{fig05}(a) demonstrates a fairly smooth dependence, which differs only little in shape from the corresponding plane-wave distribution. This, in particular, is an obvious consequence of the absence of any resonant $\pi\eta$ interaction in the region under study. At low energies (lower panels), the overall increase of the reaction yield
due to $\eta d$ attraction
results in a noticeable increase of the plane-wave cross section, without a significant change of its shape. On the whole, as we see, the situation is similar to that noted above for the total cross section. Namely, the impulse approximation visibly underestimates the data, while inclusion of interaction in the final state makes it possible to substantially improve the agreement.
The calculations by Egorov underestimate the data in the high-mass region.
As discussed later,  some features originating from the elementary amplitudes
do not appear in mass distributions of his calculations in contrast to the calculations by Egorov and Fix
and to the experimental data.
There exists some problem in the treatment of the Born terms in his
calculation.

A similar picture is observed in the case of the $d\sigma/dM_{\pi d}$ distribution
shown in Fig.~\ref{fig05}(b). Here the shape of the plane-wave cross section generally agrees with the data. A noticeable difference is seen only in the bin  $E_\gamma=0.85-0.94$ GeV, where the theoretical cross section is slightly shifted to the higher mass.
At higher energies, the distribution $d\sigma/dM_{\pi d}$ exhibits a clearly visible maximum at $M_{\pi d}\approx M_N+M_\Delta$, which apparently corresponds to the formation of $\Delta(1232)$.
As one can see from the figure, although the position of the maximum is predicted fairly well by the
theoretical calculations, there is a visible underestimation of the experimental cross section at the edges of the peak, especially in the lower part of the spectrum. The reason for this discrepancy is not very clear. It is quite possible that the approximation used in Ref.~\cite{egor13}, in which the $\pi^0d$ interaction is reduced to absorption, and is described by a trivial attenuation factor, turns out to be too rough, and more refined effects have to be taken into account. In particular, it is well known that the three-body $\pi NN$ calculations predict an $N\Delta$ resonance with $I(J^\pi)=0(2^+)$, having a mass approximately equal to $M_N+M_\Delta$ (see, e.g., Refs.~\cite{GalGarc,FixKol}). This resonance might influence the reaction dynamics, and its inclusion might improve the agreement. We will turn to this issue again in Sec.~\ref{sec:dis}.

Perhaps, the most interesting is the distribution $d\sigma/dM_{\eta d}$  shown
in Fig.~\ref{fig06}(a).
Here, the $S$-wave attraction between the $\eta$ meson and the deuteron leads to a rapid rise of the cross section in the region of low relative momenta
in the $\eta d$ system. However, although the data generally demonstrate this trend, the sharpness of the maximum in the experimental cross section at $M_{\eta d}\approx M_\eta+M_d$ turns out to be less pronounced. This is especially evident in the high energy region $E_\gamma =1.01$--$1.15$ GeV. Here, the agreement with the plane-wave cross section is even somewhat better than when the interaction is taken into account.

As for the theoretical predictions of Ref.~\cite{egor20} (solid black curves in Figs.~\ref{fig05} and \ref{fig06}), they strongly contradict our data. What is important, even the most characteristic features of the spectra, such as the $\Delta$ maximum in the $d\sigma/dM_{\pi d}$ distribution and the rapid rise of $d\sigma/dM_{\eta d}$ at $M_{\eta d}\to M_\eta +M_d$, are not reproduced by this model. The reasons for such a principle disagreement are obviously difficult to understand without a detailed analysis of the individual mechanisms included in this calculation.

It is also worthwhile to note a significant difference between the
theoretical calculations and our data for the deuteron angular distribution $d\sigma/d\Omega_d$ in Fig.\,\ref{fig06}(b). The model in Ref.~\cite{egor13} predicts a rather sharp maximum at backward angles. Such a strong angular dependence is due to the coherent nature of the reaction mechanism. As a consequence, the cross section is basically governed by the square of the deuteron scalar form factor, which falls rapidly with increasing momentum transfer. This, in turn, leads to a rapid decrease of the cross section, so that its major fraction is located in a rather narrow region of the very backward deuteron angles $\theta_d$. With increasing photon energy, the range of the available momentum transfer also increases, leading to stronger sharpening of the cross section at the maximum.

As can be seen from the figure, this simple picture is generally not confirmed by the measurements. In contrast to the
theoretical calculation, the observed cross section demonstrates a fairly smooth dependence on $\cos\theta_d$, so that in the region of the forward deuteron angles, the difference reaches one order of magnitude. Furthermore, there is no sign in the data for the sharp maximum for the backward going deuterons, as predicted by the  model.
The calculations by Egorov do not show a peak at $\cos\theta_d=-1$.
However, they also strongly underestimate
the data at forward angles.

This result is somewhat unexpected, since, as mentioned above, the main factor determining $d\sigma/d\Omega_d$ is the deuteron form factor. Other components such as elementary amplitude are of secondary importance here. In this regard, this angular distribution is expected to have weak model dependence, so that the use of a sophisticated deuteron wave function should be sufficient for a realistic description of this observable. This notion is however in qualitative disagreement with the data. It is especially surprising that the difference turns out to be so large. It looks as if there is some additional very important mechanism that allows the large transferred momentum to be shared between the nucleons, and thus increases the cross section at forward deuteron angles. To understand the situation better, further research in this area is needed.

\subsection{Double-differential cross sections}\label{sec:ddcs}
\begin{figure*}
\begin{center}
\includegraphics[width=0.95\textwidth]{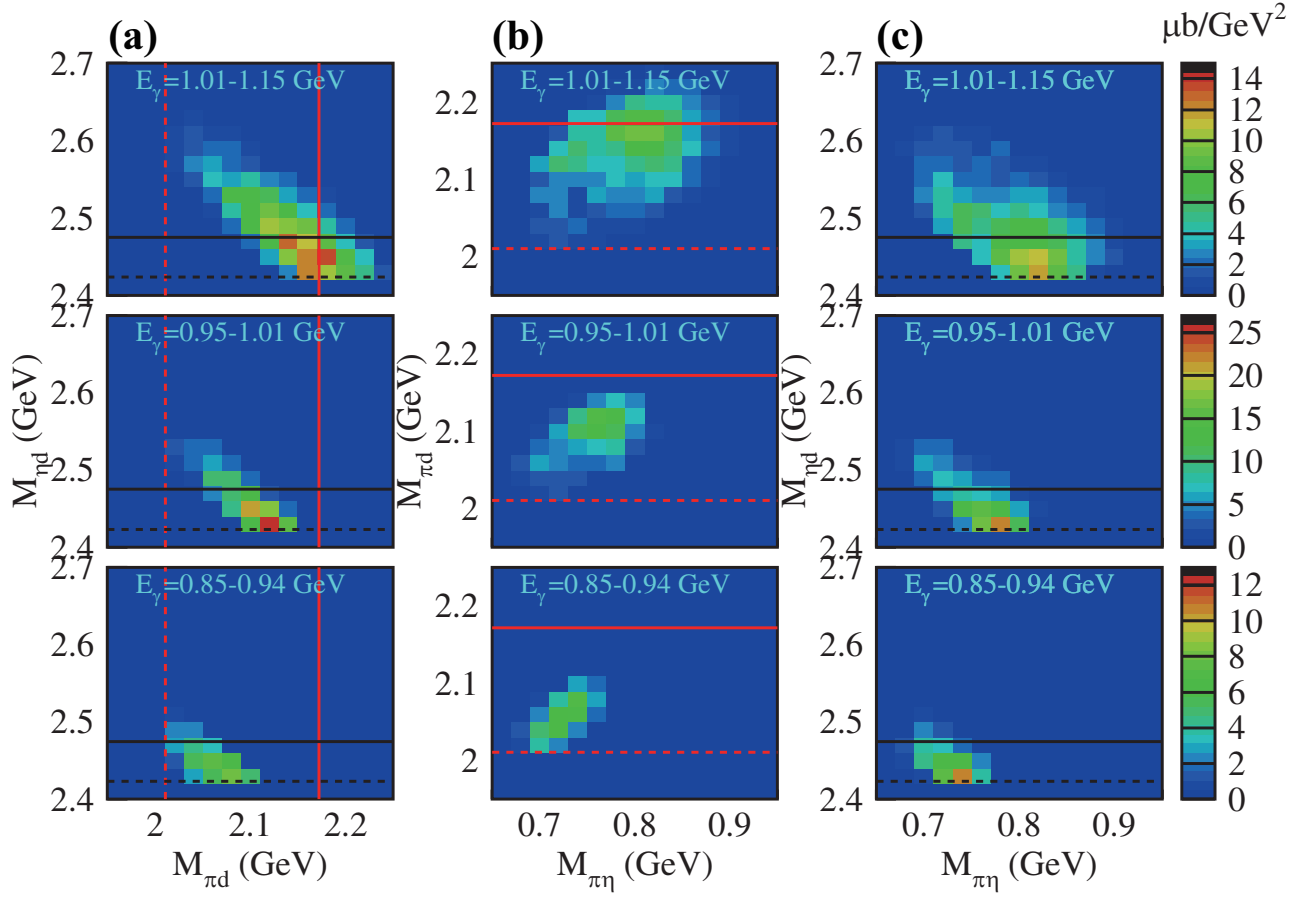}
\end{center}
\caption{Double-differential cross sections
$d^2\sigma/dM_{\eta d}/d M_{\pi d}$ (a),
$d^2\sigma/dM_{\pi d}/d M_{\pi \eta}$ (b), and
$d^2\sigma/dM_{\eta d}/d M_{\pi \eta}$ (c)
for $E_\gamma=0.85$--0.94 GeV (bottom), 0.95--1.01 GeV (center),
and 1.01--1.15 GeV (top).
The red solid (dashed) lines represent $M_{\pi d}=M_N+M_\Delta$ ($M_{\pi^0}+M_d$),
and the black solid (dashed) indicate $M_{\eta d}=M_N+{M_{N^*}}$ ($M_\eta+M_d$)
where $N^*$ stands for $N(1535)1/2^-$.
}\label{fig07}
\end{figure*}

To investigate correlation between two of the invariant masses
$M_{\pi \eta}$, $M_{\eta d}$, and $M_{\pi d}$,
we also deduced the double-differential cross sections
$d^2\sigma/dM_{\eta d}/d M_{\pi d}$,
$d^2\sigma/dM_{\pi d}/d M_{\pi \eta}$, and
$d^2\sigma/dM_{\eta d}/d M_{\pi \eta}$.
A double-differential cross section is given by
\begin{widetext}
\begin{equation}
\frac{d^2\sigma}{dV_i\, dV_j}(E_\gamma,V_i,V_j) = \frac{
\mathcal{N}_{\pi\eta d}(E_\gamma,V_i,V_j)
}{
\mathcal{N}'_\gamma\,
\mathcal{N}_\tau\,
\mathcal{A}(E_\gamma,V_i,V_j)\,
\mathcal{B}({\pi^0\to \gamma\gamma})\,
\mathcal{B}({\eta\to \gamma\gamma})\,
\Delta V_i\,
\Delta V_j
\label{eq:dcs2}
}\end{equation}
\end{widetext}
where $V_i$ and $V_j$ stand for two independent
invariant masses from  $M_{\pi\eta}$, $M_{\eta{d}}$, and $M_{\pi{d}}$.
Figures~\ref{fig07}(a), (b), and (c) show the obtained double-differential cross sections
$d^2\sigma/dM_{\eta d}/d M_{\pi d}$,
$d^2\sigma/dM_{\pi d}/d M_{\pi \eta}$, and
$d^2\sigma/dM_{\eta d}/d M_{\pi \eta}$, respectively,
together with
the red solid (dashed) lines representing $M_{\pi d}=M_N+M_\Delta$ ($M_{\pi^0}+M_d$),
and the black solid (dashed) indicating $M_{\eta d}=M_N+{M_{N^*}}$ ($M_\eta+M_d$)
where $N^*$ stands for $N(1535)1/2^-$.

The correlation between $M_{\eta{d}}$ and $M_{\pi{d}}$ at $E_\gamma=1.01$--1.15 GeV
 in Fig,~\ref{fig07}(a)
shows an enhancement near the point
 $(M_{\pi d}, M_{\eta d}) = (M_N+M_\Delta, M_\eta + M_d)$.
Whether or not the enhancement
at $E_\gamma= 0.85$--0.94 and 0.95--1.01 GeV
actually shows up is unclear,
since the phase space does not cover
$M_{\pi d}=M_N+M_\Delta$.
An enhancement near the $\eta d$ threshold is seen in Fig.~\ref{fig07}(c)
at all the incident energies in the $\left(M_{\eta{d}},M_{\pi{\eta}}\right)$ plot.
Additionally,
an enhancement at $M_{\pi d}=M_N+M_\Delta$ is observed
at $E_\gamma = 1.01$--1.15 GeV in the
$\left(M_{\pi{d}},M_{\pi{\eta}}\right)$ plot
as shown in Fig.~\ref{fig07}(b).
Because of the limited phase space,
it is difficult to distinguish the two conditions
$M_{\pi d} \approx M_N+N_\Delta$ and $M_{\eta d} \approx M_\eta+M_d$.
However, the lower edge of the peak near the $\eta d$ threshold does not visibly change, regardless of incident energies,
and the peak becomes wider at higher incident energies.
The resonance-like structure in the $\pi^0d$ channel,
which is clearly observed
near $M_{\pi d}=M_N+M_\Delta$ at the highest incident energies, is the main source of broadening of the maximum in the $\eta d$ spectrum,
resulting from the strong $\eta d$ attraction.

\section{Phenomenological analysis}\label{sec:dis}

\begin{figure*}
\begin{center}
\includegraphics[width=0.7\textwidth]{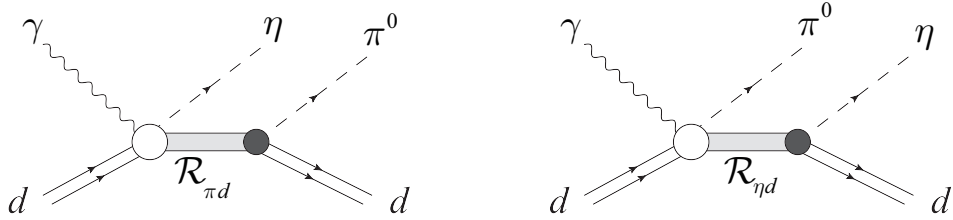}
\end{center}
\caption{Schematic representation of the $\gamma d\to\pi^0\eta d$ amplitude (\ref{eq:1c}) used in our phenomenological analysis.}
\label{fig11}
\end{figure*}

As discussed in the previous section, the microscopic models of
Refs.~\cite{egor13,egor20} are unable to explain some important properties of the $\gamma d\to \pi^0\eta d$ reaction observed in the experiment. In this regard, for further analysis, we use a purely phenomenological model, in which the amplitude consists of two terms
\begin{equation}\label{eq:1c}
T=T_{\pi(\eta d)}+T_{\eta(\pi d)}\,,
\end{equation}
schematically illustrated in Fig.\,\ref{fig11}. Our method follows the same lines as in Ref.\,\cite{ishi21}, where similar questions were addressed for the $NN(1535)1/2^-$ interaction.

For the intermediate quasi-two-body states $\mathcal{R}_{\eta d}$ and
$\mathcal{R}_{\pi d}$ we adopt the pole approximation. In particular, in $\pi^0 d$ we took into account the previously mentioned pole, predicted by Faddeev calculations of the $\pi NN$ system in the $I(J^\pi)=1(2^+)$ state \cite{GalGarc,FixKol}, sometimes treated in the literature as the $N\Delta$ dibaryon resonance
$\mathcal{D}_{12}(2150)$~\cite{GalGarc}. This configuration corresponds to the $L_J=P_2$ state in the $\pi d$ system. The pole position predicted by the
theoretical calculations
 is approximately at $2.10+i\,0.14$ GeV. The corresponding $\pi d$ scattering phase shift $\delta_{P_2}$ demonstrates purely resonant behavior, which is accompanied by a rapid decrease of the inelasticity parameter \cite{FixKol}. The other phase shifts are rather small in the energy region considered and can be neglected. Therefore, for $\mathcal{R}_{\pi d}$, we use the pure Breit-Wigner ansatz in the $1(2^+)$ partial wave, in which the mass $M$ and the width $\Gamma$ are treated as adjustable parameters.

The $\eta d$ pole in the $I(J^\pi)=0(1^-)$ configuration is located
near the $\eta d$ threshold energy. The results of various analyses
\cite{gree97,Fix:2000hf,Barnea:2015lia} indicate that this pole is
on the unphysical sheet so that it generates a virtual $\eta d$ state.
For $\mathcal{R}_{\eta d}$, the Flatt\'e parametrization~\cite{flat76,kala09}
was adopted in Ref.~\cite{ishi21}, allowing us to take into account
the closeness of the virtual $\eta d$ level to the threshold energy.
To deduce the $\eta d$ scattering parameters $a_{\eta d}$ and $r_{\eta d}$,
we replaced the Flatt\'e ansatz by the low-energy formula
\begin{equation}
\mathcal{R}_{\eta d}\sim \left({\frac{1}{a_{\eta d}} + \frac{1}{2} r_{\eta d}\, p_\eta^2 -i {p_\eta}}\right)^{-1}\,,
\end{equation}
where
$p_\eta$ is as previously the $\eta$ momentum in the $\eta d$ CM frame.
%The corresponding expression for the total width reads
%$\Gamma= \Gamma_0 + g p_\eta c$,
%where $\Gamma_0$ is the effective constant width associated with the opened %channels, coupled to $\eta d$, that is, $\pi NN$, $\pi\pi NN$, and $\pi\pi d$. The %constant
%$g$ determines the coupling of $\mathcal{R}_{\eta d}$ to the $\eta d$ state,
%and $p_\eta$ is as previously the $\eta$ momentum in the $\eta d$ CM frame.
For completeness, in addition to the dominant $L=0$ decay we also allow the $L=2$ transition of $\mathcal{R}_{\eta d}$ to the final $\eta d$ state.

The simultaneous fit was performed to the
$d\sigma/dM_{\eta d}$ and $d\sigma/dM_{\pi d}$ data
at $E_\gamma=1.01$--1.15 GeV and $E_\gamma=0.95$--1.01 GeV. The fitted functions are built according to (see also Eqs.\,(1) to (4) in Ref.\,\cite{ishi21})
\begin{equation}\label{eq:1d}
\frac{d\sigma}{dM_{\eta d}}=\alpha_0\int A(M_{\eta d},M_{\pi d})
\,V_{\rm PS}(M_{\eta d},M_{\pi d})\,dM_{\pi d}\,,
\end{equation}
where $V_{\rm PS}(M_{\eta d},M_{\pi d})$ is the reaction phase space and $A(M_{\eta d},M_{\pi d})$ is given by
\begin{widetext}
\begin{equation}\label{eq:1e}
\displaystyle
A(M_{\eta d},M_{\pi d}) =
\left(1+\alpha_2\,p_\eta^4
\right) \left|\mathcal{R}_{\eta d}(M_{\eta d})
\vphantom{\sum}
\right|^2 + \alpha_1\,\left|\mathcal{R}_{\pi d}(M_{\pi d})
\vphantom{\sum}
\right|^2\,.
\end{equation}
\end{widetext}
Here, the experimental mass resolutions
are convoluted in the evaluation of Eq.~(\ref{eq:1d}).
Similarly to Ref.~\cite{ishi21}, we neglect in Eq.\,(\ref{eq:1e}) the interference of the $\gamma d\to \pi^0 \mathcal{R}_{\eta d}\to \pi^0\eta d$ and $\gamma d\to \eta \mathcal{R}_{\pi d}\to \pi^0\eta d$ transitions. This approximation may not be
crucial since the angular momentum of $\eta$ emission differs
between the corresponding mechanisms~\cite{ishi21}.
For the $d\sigma/dM_{\pi d}$ distribution, a similar expression was used, where the integration is carried out, of course, over $M_{\eta d}$.

\begin{figure*}
\begin{center}
\includegraphics[width=0.8\textwidth]{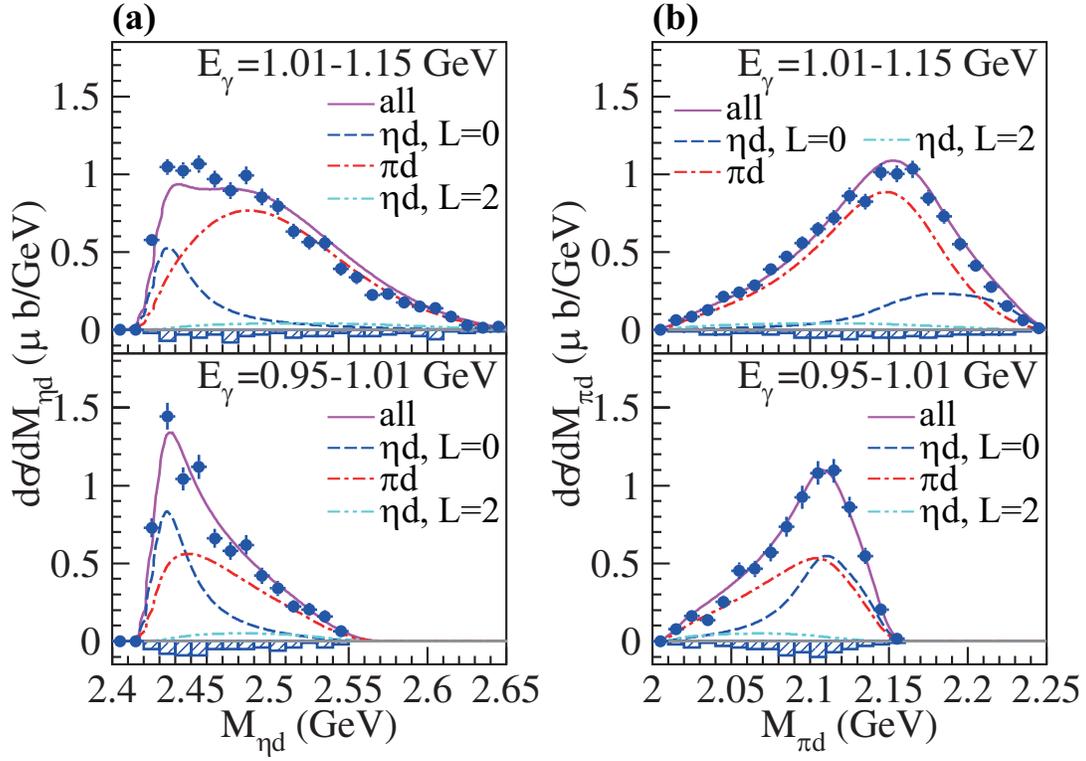}
\end{center}
\caption{
Differential cross sections $d\sigma/dM_{\eta d}$
(a)
and $d\sigma/dM_{\pi d}$ (b) together with
the fitted functions (magenta solid curves)
at $E_\gamma=0.95$--1.01 GeV (bottom) and 1.01--1.15 GeV (top).
The blue dashed and cyan double-dotted curves
show the $S$-  and $D$-wave decay contributions of $\mathcal{R}_{\eta d}$,
respectively.
The red dot-dashed curves represent the contribution from the second term in Eq.\,(\ref{eq:1e}).}
\label{fig09}
\end{figure*}

Our solution is presented in Fig.~\ref{fig09}, where the contributions determined by the two terms in Eq.\,(\ref{eq:1e})
are separately shown. The resulting $\chi^2$ value is 131.3 for 76 data points. The $\pi^0 d$ parameters coming out of the fit are
$M=2.158_{-0.003}^{+0.003}$ GeV and $\Gamma=0.116_{-0.007}^{+0.004}$ GeV, respectively. They are consistent with the results of the mass fit with
Flatt\'e parametrization in Ref.~\cite{ishi21}. Furthermore, they are quite close to $M=2.147$, $\Gamma=0.12$ and $M=2.091$, $\Gamma=0.16$ obtained for the $N\Delta$ $\mathcal{D}_{12}$ resonance within the three-body calculations in Refs.\,\cite{GalGarc} and \cite{FixKol}, respectively.

Adjusting the model parameters to the $d\sigma/dM_{\eta d}$ and $d\sigma/dM_{\pi d}$ data at $E_\gamma=0.95$--1.01 and 1.01--1.15 GeV gives
$a_{\eta d}=\pm \left(0.7_{-0.6}^{+0.8}\right)+i\left(0.0_{-0.0}^{+1.5}\right)$ fm
and
$r_{\eta d}=\mp \left(4.3_{-2.9}^{+8.6}\right)-i\left(6.7_{-8.4}^{+6.0}\right)$ fm.
Here, the double sign in $r_{\eta d}$ corresponds to that in $a_{\eta d}$.
It cannot be uniquely determined from the fit. However, for qualitative agreement with the theoretical results \cite{gree97,Fix:2000hf,Barnea:2015lia}
which support only a virtual $\eta d$ level, a plus sign of $a_{\eta d}$ must be taken.
The $d\sigma/dM_{\eta d}$ and $d\sigma/dM_{\pi d}$ obtained
in this work are rather insensitive to $r_{\eta d}$ due to
the limited phase space at $E_\gamma < 1.15$ GeV,
making the statistical error of $r_{\eta d}$ quite large.
These large uncertainties also make it difficult to determine $a_{\eta d}$ more or less precisely.
The value of $a_{\eta d}$ thus obtained is in reasonable agreement with the three-body results. In particular, it is quite close to $a_{\eta d}= 1.23+i\,1.11$ fm obtained in Ref.~\cite{egor13} with a set of $\eta N$ parameters
($a_{\eta N}=0.50+i\,0.33$ fm)
adjusted to the $\eta N$ scattering amplitude from Ref.~\cite{wilk93}.

%\subsection{Angular distribution of deuteron emission}

\section{Summary}\label{sec:sum}
The $\gamma{d}${$\to$}$\pi^0\eta{d}$ reaction has been experimentally
studied at $E_\gamma < 1.15$ GeV. We have measured the total cross section $\sigma$ as a function of $E_\gamma$, which demonstrates a rapid rise in the region from $E_\gamma=0.8$ up to 1 GeV. The existing theoretical calculations incorporating microscopic treatment of final state interaction reproduce
$\sigma(E_\gamma)$ rather well. As our analysis shows, taking into account the interaction effect  noticeably influences the result and is of decisive importance for the observed fairly good agreement between
theoretical calculation and experimental data.

We have also measured the differential cross sections
$d\sigma/dM_{\pi \eta}$, $d\sigma/dM_{\pi d}$,
$d\sigma/dM_{\eta d}$, and $d\sigma/d\Omega_d$ for the first time.
The $\eta d$ interaction is clearly seen in the $\eta d$ invariant mass spectrum $d\sigma/dM_{\eta d}$. The corresponding enhancement due to the large scattering length $a_{\eta d}$ is qualitatively reproduced by the theoretical calculation, although it predicts a too sharp maximum at the $\eta d$ threshold. The difference between the
calculation
and the experimental results for $d\sigma/dM_{\eta d}$ becomes especially noticeable with increasing energy.

The $\eta d$ scattering parameters obtained from the phenomenological analysis of $d\sigma/dM_{\eta d}$ and $d\sigma/dM_{\pi d}$ are
$a_{\eta d}=\pm \left(0.7_{-0.6}^{+0.8}\right)+i\left(0.00_{-0.0}^{+1.5}\right)$ fm
and
$r_{\eta d}=\mp \left(4.3_{-2.9}^{+8.6}\right)-i\left(6.7_{-8.4}^{+6.0}\right)$ fm,
where the double sign in $r_{\eta d}$ corresponds to that in $a_{\eta d}$.
Because of the uncertain sign of ${\rm Re}[a_{\eta d}]$,
it is not clear whether a bound or a virtual $\eta d$ state is generated. The existing microscopic theories generally support the idea that the strength of the $\eta d$ interaction is not sufficient to bind this system, so that the observed increase should be associated with a virtual $\eta d$ level.

The model predictions for the angular distribution $d\sigma/d\Omega_d$ exhibit a pronounced maximum at the very backward deuteron angles. As discussed in Sect.\,\ref{sec:dcs} such dependence is mainly governed by the strong dependence of the deuteron form factor on the momentum transfer.
At the same time, the data show no sign for any sharp peaking for backward-going deuterons. As a result, in the forward deuteron hemisphere
the data are underestimated by almost a factor of ten.

The sharp angular dependence, predicted by the
theoretical calculations, might be softened by different second order processes, in which the spectator nucleon is also involved and thus the transferred momentum is shared between the both nucleons. However, direct calculations show that the effect of such mechanisms is limited and does not allow one to explain the observed difference. Since the discrepancies in the deuteron angular distribution are rather serious, they give reason to doubt the validity of our general notion about the main mechanisms of the $\gamma d\to\pi^0\eta d$ reaction (impulse approximation plus interaction in the final state). Therefore, further research in this area is highly desirable.

\begin{acknowledgments}
The authors thank the ELPH accelerator staff for stable operation of
the accelerators to provide the primary electron beam
with good quality during the long-term FOREST experiments.
They acknowledge Mr.\ Kazue~Matsuda, Mr.~Ken'ichi~Nanbu, and Mr.~Ikuro~Nagasawa for their technical support.
They also express their gratitude to Prof.\ Mikhail Egorov for the theoretical calculations of the total and differential cross sections.
They are grateful to Prof.\ Bernd Krusche for giving us the numerical values of the total cross sections
measured at the MAMI facility.
This work was partly supported by the Ministry of Education, Culture, Sports, Science
and Technology, Japan (MEXT) and Japan Society for the Promotion of Science (JSPS)
through Grants-in-Aid
for Specially Promoted Research No.\ 19002003,
for Scientific Research (A) Nos.\ 24244022 and  16H02188,
for Scientific Research (B) Nos.\ 17340063 and 19H01902,
for Scientific Research (C) No.\ 26400287, and
for Scientific Research on Innovative Areas Nos.\ 19H05141,
19H05181, and 21H00114.
The model calculations
were
supported by the Russian Science Foundation, Grant No. 22-42-04401.
\end{acknowledgments}

\end{document}